%% file: main.tex
\documentclass[conference]{IEEEtran}
\IEEEoverridecommandlockouts

\usepackage{amsmath,amssymb,amsfonts,bm}
\usepackage{graphicx}
\usepackage{subcaption}
\usepackage{cite}
\usepackage{siunitx}
\usepackage{algorithm}
\usepackage{algpseudocode}
\usepackage{standalone}   
\usepackage{pgfplots}

\usepackage{amsthm}
\theoremstyle{remark}
\newtheorem{remark}{Remark}

\usepackage{placeins}  

\tikzset{every picture/.style={/tikz/font=\footnotesize}}

\pgfplotsset{compat=newest}

\newcommand{\algcond}[1]{\parbox[t]{\dimexpr\linewidth-\algorithmicindent\relax}{#1}}

\usetikzlibrary{positioning}
\usetikzlibrary{intersections}

\usepackage{tikz}
\usepackage[dvipsnames]{xcolor}
\usepackage{standalone}
\usetikzlibrary{arrows.meta,calc,positioning,decorations.pathreplacing}
\definecolor{tileBlue}{RGB}{150,180,220}
\definecolor{panelEdge}{RGB}{30,30,30}
\definecolor{dashBlue}{RGB}{45,95,160}
\definecolor{phonePurple}{RGB}{168,100,170}

\usepackage{tikz}
\tikzset{
  pics/phone/.style={code={
    \fill[black,rounded corners=0.5mm] (0,0) rectangle (0.6,1.1);
    \fill[rounded corners=0.5mm, white!90!black] (0.05,0.15) rectangle (0.55,0.95);
    \draw[white, thick] (0.22,1.025) -- (0.38,1.025);
    \fill[white](0.3,0.075) circle (.4ex);
  }}
}
            
\title{
Exploiting Near-Field Dynamics with Movable Antennas to Enhance Discrete Transmissive RIS
\thanks{This work was supported in part by the German Federal Ministry of Research Technology and Space (BMFTR) in the course of the 6GEM+ Transfer Hub under grant 16KIS2411.}
}

\author{
\IEEEauthorblockN{Marjan Boloori, Chu Li, Aydin Sezgin}
\IEEEauthorblockA{Department of Digital Communication Systems, Ruhr University Bochum, Germany \\
Email: \{marjan.boloori, chu.li, aydin.sezgin\}@rub.de}
}

\begin{document}
\maketitle

\begin{abstract} 
The design of low-complexity transceivers is crucial for the deployment of next-generation wireless systems. In this work, we combine two emerging concepts, movable antennas (MA) and transmissive reconfigurable intelligent surfaces (TRIS), which have recently attracted significant attention for enhancing wireless communication performance. In particular, we propose a compact base station (BS) architecture that integrates a single MA with a TRIS operating in their near-field region. We address the joint optimization of the MA location and the quantized TRIS phase configuration. Due to the non-convex coupling between spatial positioning and discrete phase constraints, an alternating optimization (AO) framework is developed, where the MA position is updated via gradient ascent (GA) and the TRIS phases are optimized through quantized phase alignment. Simulation results demonstrate that the proposed architecture significantly outperforms conventional BS designs equipped with fixed fully-active antenna arrays under the same channel model and transmit power constraint. Moreover, MA repositioning effectively mitigates the performance degradation caused by discrete TRIS phase quantization in near-field propagation environments. This reveals a favorable trade-off between hardware complexity and spatial signal processing, where the spatial adaptability of the MA can compensate for low-resolution TRIS phase control.
\end{abstract}

\begin{IEEEkeywords}
Transmissive RIS, movable antenna, near-field, discrete phase-shift quantization, cascaded path-loss.
\end{IEEEkeywords}

\section{Introduction}
Reconfigurable intelligent surfaces (RIS) have emerged as a promising technology for sixth-generation (6G) wireless systems due to their capability to control the wireless propagation environment and enhance communication performance~\cite{wu2019towards, KevinResilience}. 
An RIS consists of a large number of passive elements that manipulate the phase and amplitude of incident electromagnetic waves, enabling constructive signal combination at desired receivers and interference suppression~\cite{ChuJ}. Owing to this operation, RIS avoid the additional noise and self-interference typically associated with conventional relay technologies while maintaining low power consumption and implementation cost~\cite{yan2020passive}. 

RIS architectures can generally be categorized into reflective~\cite{li2021joint, chen2022irs} and transmissive~\cite{zeng2021reconfigurable, 10886969} configurations. In reflective RIS systems, both the transmitter and the receiver are located on the same side of the surface, which may introduce undesired coupling and deployment constraints. In contrast, transmissive RIS (TRIS) places the transmitter and receiver on opposite sides of the surface, enabling more flexible propagation control and additional spatial degrees of freedom for channel shaping~\cite{li2023toward}.

Unlike conventional wireless links, TRIS-assisted communication relies on a cascaded channel. The received signal propagates through the transmitter–TRIS and TRIS–receiver links, resulting in a multiplicative path-loss effect~\cite{ellingson2021path,KevinCascadedChannel}. 
This cascaded attenuation can significantly limit the achievable performance when the TRIS is located far from the transmitter. Consequently, deploying the TRIS in close proximity to the transmitter becomes an attractive architecture, as it reduces the first-hop propagation distance and enables more efficient signal shaping directly at the transmitter side. In such transmitter-oriented architectures, providing additional spatial control at the radiating element becomes particularly advantageous.

Movable antennas (MA) have recently emerged as a promising technique to realize this capability by allowing controlled repositioning of the antenna. This additional spatial flexibility enables adaptive spatial focusing and has been shown to significantly improve the received signal-to-noise ratio (SNR) under various channel conditions~\cite{ChuMarjan, zhu2023modeling}. 
However, existing MA studies mainly focus on conventional wireless systems with direct links~\cite{10354003} or reflective RIS configurations~\cite{10962171}, while the integration of MA with TRIS operating in the near-field regime remains largely unexplored.

Motivated by these insights, this work investigates the interaction between a MA and a TRIS operating in the near-field region. In this configuration, the cascaded channel implies that the MA position directly influences the effective channel gain through the element-wise propagation distances between the MA and the TRIS. By adjusting the MA location, the spatial distribution of these distances can be favorably shaped, enabling improved coherent signal combination and partially alleviating the cascaded path-loss effect. In practical implementations, TRIS hardware typically employs discrete phase shifters with limited resolution, often in the range of 1-2 bits~\cite{song2024modeling,TRIS_Practical_B5G}. Such phase quantization generally limits the achievable channel gain due to imperfect phase alignment. In this work, we show that the spatial degree of freedom introduced by MA repositioning can effectively mitigate this hardware limitation. By appropriately adapting the MA position, the system can recover a significant portion of the performance loss caused by low-resolution phase control. This observation highlights a favorable trade-off between hardware complexity and spatial signal processing, enabling a BS architecture where the spatial adaptability of the MA mitigates for the use of low-complexity TRIS hardware.

\section{System Model and Problem Formulation}

\begin{figure}[t]
\centering

    \begin{tikzpicture}[
            x=5mm,y=5mm,
            yslant=0.18,               
            scale=0.8,
            >={Latex[length=1mm]},
             every path/.style={line width=0.15pt}, 
            ray/.style={line width=0.5pt},
            label/.style={scale=0.9,inner sep=1pt}
          ]
        
          \begin{scope}[yslant=-0.18]
            \draw[dashBlue, line width=1.2pt, dashed, dash pattern=on 7pt off 5pt, rounded corners=15pt]
              (-3.7,-1.0) rectangle (10.2,11);
            \node[dashBlue, font=\small] at (3.25,-2) {Base Station};
          \end{scope}
        
          \begin{scope}[shift={(1.5,0)}]
          
          \def\W{8}
          \def\N{6}
          \def\s{0.80}     
          \def\g{0.52}     
          \pgfmathsetmacro{\m}{(\W - (\N*\s + (\N-1)*\g))/2}
          \coordinate (C) at ({\W/2},{\W/2}); 
        
          \fill[white] (0,0) rectangle (\W,\W);
          \draw[panelEdge, line width=1pt] (0,0) rectangle (\W,\W);               

          \foreach \i in {0,...,5}{
            \foreach \j in {0,...,5}{
              \pgfmathsetmacro\x{\m + \i*(\s+\g)}
              \pgfmathsetmacro\y{\m + \j*(\s+\g)}
              \fill[ tileBlue!40!white ] (\x,\y) rectangle ++(\s,\s);
              \draw[panelEdge!70, line width=0.5pt] (\x,\y) rectangle ++(\s,\s);
            }
          }

            \coordinate (RIS_bl) at (0,0);
            \coordinate (RIS_tl) at (0,\W);
            \coordinate (RIS_br) at (\W,0);
            \coordinate (RIS_tr) at (\W,\W);

            \end{scope}
            

            \begin{scope}[shift={(0,-0.2)}, yslant=0]
              \coordinate (MA) at (-2.2,7.1);
            
              \def\boxsize{1.6} 
              \draw[panelEdge, line width=0.5pt, dashed, rounded corners=3.5pt]
                ($(MA)+(-\boxsize/2,-\boxsize/2)$) rectangle
                ($(MA)+(\boxsize/2,\boxsize/2)$);
            \node[anchor=north, label] at ($(MA)+(0,2)$) {$a$};
            \node[anchor=north, label] at ($(MA)+(1.5,0)$) {$a$};
            
              \draw[panelEdge!80, line width=1.0pt] (MA) -- ++(0,-0.4);
            
                \path let \p1=(MA) in
                  coordinate (B) at (\x1-4.0,\y1+4)        
                  coordinate (D) at (\x1+4.0,\y1+4)        
                  coordinate (A) at (\x1,\y1-1.20);       
                \draw[fill=tileBlue!60, line width=0.5pt] (B) -- (D) -- (A) -- cycle;

                \path
                  ($(MA)+(-\boxsize/2,\boxsize/2)$)  coordinate (MA_tl)
                  ($(MA)+(\boxsize/2,\boxsize/2)$)   coordinate (MA_tr)
                  ($(MA)+(-\boxsize/2,-\boxsize/2)$) coordinate (MA_bl)
                  ($(MA)+(\boxsize/2,-\boxsize/2)$)  coordinate (MA_br);
                \draw[panelEdge, line width=0.5pt, dashed, rounded corners=3.5pt]
                  (MA_bl) rectangle (MA_tr);

            \draw[panelEdge!80, <->, line width=0.55pt]
              ($(MA)+(-1,1)$) -- ($(MA)+(1,1)$);
            
            \draw[panelEdge!80, <->, line width=0.55pt]
              ($(MA)+(1,-0.85)$) -- ($(MA)+(1,1)$);
        
              \node[anchor=north, label] at ($(MA)+(0.2,-1.3)$) {Movable};
              \node[anchor=north, label] at ($(MA)+(0.2,-1.9)$) {Antenna};
            \end{scope}

            \draw[panelEdge!25, line width=0.25pt] (MA_tl) -- (RIS_tl);
            \draw[panelEdge!25, line width=0.25pt] (MA_tr) -- (RIS_tr);
            \draw[panelEdge!25, line width=0.25pt] (MA_bl) -- (RIS_bl);
            \draw[panelEdge!25, line width=0.25pt] (MA_br) -- (RIS_br);
        
            \node[transform shape=true] at (16.5,-1.1) {%
              \tikz[baseline]{
                \pic[scale=1.6] {phone}; 
              }%
            };
            \node[below=12pt, font=\small] at (16.5, -1.1) {User};

            \draw[ray, -{Latex[length=2.6mm]}]
          (C) -- (16,-0.9)
          node[pos=0.75, below=-10pt, label] {$g$};
        
            \fill[panelEdge] (C) circle (0.06);
        
          \node[anchor=north west, label] at (3.5,-0.1) {Transmissive RIS};

          \draw[ panelEdge!80, line width=0.7pt, -{Latex[length=2.6mm]}]
          (-2.2,6.5) -- (C)
          node[pos=0.5, above right=1.5pt, label, black] {$f(\boldsymbol{t})$};
          
            
            
            

    \end{tikzpicture}
    
    \caption{System model, the suggested BS design with TRIS and MA, and the user device at the far-field.}
    \label{fig:SysModel}
\end{figure}
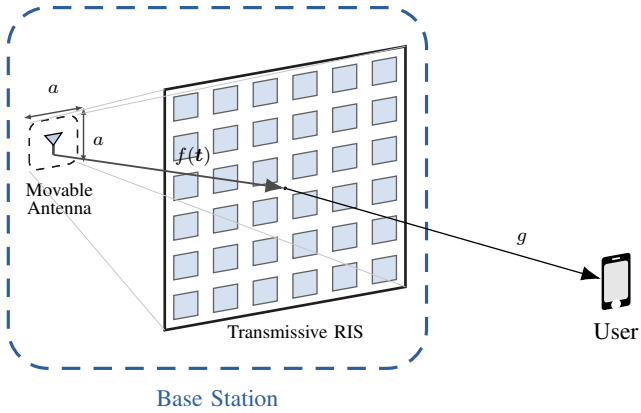

\subsection{System Model}
As shown in Fig.~\ref{fig:SysModel}, we consider a BS architecture consisting of a single MA and a TRIS. 
The MA is located at position $\boldsymbol{t} = [x_t, y_t, z_t]^T$ and can be flexibly repositioned within a two-dimensional region denoted by $\mathcal{R}_{\text{MA}} = a \times a \,\mathrm{m^2}$.

The TRIS comprises $N$ number of transmissive elements arranged on a planar surface. 
The Cartesian coordinate of the $n$-th TRIS element is denoted by $\boldsymbol{e}_n = [x_n, y_n, 0]^T$, where $n \in \mathcal{N} \triangleq \{1,2,\dots,N\}$. 
The inter-element spacing between adjacent TRIS elements is denoted by $d_\text{TRIS}$.

The quantized phase-shift configuration of the TRIS is represented by
\begin{equation}
    \tilde{\boldsymbol{\phi}} = [\tilde{\phi}_1, \tilde{\phi}_2, \dots, \tilde{\phi}_n, \dots, \tilde{\phi}_N]^T ,
\end{equation}
where $\tilde{\phi}_n$ denotes the $b$-bit quantized phase shift assigned to the $n$-th transmissive element. Each discrete phase $\tilde{\phi}_n$ is obtained by mapping the corresponding continuous optimal phase $\phi_n$ to the nearest value in a predefined quantization set
\begin{align}
    \boldsymbol{\Phi}_b = \big \{  0, \frac{1}{2^{b-1}}\pi, \dots, \frac{2^b-1}{2^{b-1}}\pi  \big \} .   
\end{align}
where $b$ denotes the number of quantization bits.
The TRIS is deployed in the near-field region of the MA and is assumed to be fully transmissive, meaning that the incident electromagnetic wave is transmitted through the surface without reflection.
The near-field boundary, given by the Rayleigh distance, can be calculated as
\begin{equation}
    D_\text{Ray} = 2D_\text{Aper}^2/\lambda,
\end{equation}
where $\lambda$ represents the wavelength and $D_\text{Aper}$ denotes the largest dimension of the TRIS~\cite{RISNearField}. In this context, $D_\text{Aper} = \sqrt{2N}d_\text{TRIS}$. This distance represents the spatial separation between the MA region and the TRIS. Subsequently, for each TRIS size, the relative Rayleigh distance for the MA is evaluated to ensure that the MA and TRIS remain within the near-field region. Moreover, the user is located on the opposite side of the TRIS in the far-field. 

Let $f_{n}(\boldsymbol{t})$ denote the near-field channel coefficient between the MA and the $n$-th element of the TRIS which is a function of the MA position vector $\boldsymbol{t}$. The corresponding propagation distance between the transmitter (MA) and the TRIS is therfore given by
\begin{equation}
    d_{n}^{\text{T}}(\boldsymbol{t}) 
    = \left\| \boldsymbol{t} - \boldsymbol{e}_{n} \right\|,
\end{equation}
where $\boldsymbol{e}_n$ represents the Cartesian coordinates of the $n$-th TRIS element. 
Under a spherical-wave propagation model, the resulting channel coefficient can be expressed as
\begin{equation}\label{eq:f}
    f_{n}(\boldsymbol{t}) 
    = \sqrt{\frac{\lambda G_f F_f}{4\pi}} 
    \frac{1}{d_{n}^{\text{T}}(\boldsymbol{t})}
    e^{-j\frac{2\pi}{\lambda} d_{n}^{\text{T}}(\boldsymbol{t})},
\end{equation}
where $\lambda$ denotes the wavelength, and $G_f$ and $F_f$ represent the antenna gain of the MA and the normalized power radiation pattern of the TRIS side facing the MA, respectively \cite{channelmodelFG}.
Similarly, the propagation distance between the $n$-th TRIS element and the receiver located at $\boldsymbol{u}$ is given by
\begin{equation}
    d_{n}^{\text{R}} = \left\| \boldsymbol{u} - \boldsymbol{e}_n \right\|.
\end{equation}
Accordingly, the corresponding channel coefficient is expressed as
\begin{equation}\label{eq:g}
    g_{n} 
    = \sqrt{\frac{\lambda G_g F_g}{4\pi}} 
    \frac{1}{d_{n}^{\text{R}}}
    e^{-j\frac{2\pi}{\lambda} d_{n}^{\text{R}}},
\end{equation}
where $G_g$ and $F_g$ denote the antenna gain and the normalized power radiation pattern of the TRIS side facing the user, respectively.

Based on (\ref{eq:f}) and (\ref{eq:g}), the received signal by the user can be modeled as
\begin{align}\label{eq:ReceivedSignal}
    & y = \sqrt{P} \sum_{n = 1}^{N} f_{n} g_{n} \Gamma e^{j\tilde{\boldsymbol{\phi}}_{n}} s + n_0, 
\end{align}
where $P$ represents the transmit power, $n_0 \sim \mathcal{CN}(0, \sigma ^2)$ denotes the additive Gaussian noise with variance $\sigma ^2$, $s$ is the transmit signal, $\Gamma \in [0,1]$ is the transmission loss of the TRIS. 
Therefore, the corresponding received SNR is given by
\begin{equation}\label{eq:SNRFirst}
    \text{SNR}(\boldsymbol{t},\tilde{\boldsymbol{\phi}}) = \kappa \left| c(\boldsymbol{t},\tilde{\boldsymbol{\phi}}) \right|^2,
\end{equation}
where
\begin{equation}
    \kappa = \frac{\lambda ^2 \Gamma ^2 P G F}{16 \pi^2 \sigma^2},
\end{equation}
with $G = G_fG_g$, and $F = F_fF_g$.
The function $c(\boldsymbol{t},\tilde{\boldsymbol{\phi}})$ is defined as
\begin{equation}\label{eq:c}
    c(\boldsymbol{t},\tilde{\boldsymbol{\phi}}) = \sum_{n=1}^{N} \frac{1}{d^T_n(\boldsymbol{t}) d_n^R} e{^{j \psi_n(\boldsymbol{t})}},
\end{equation}
with ${\psi_n(\boldsymbol{t})} = {   \tilde{\phi}_n - \frac{2\pi}{\lambda} ( d^T_n(\boldsymbol{t})+d_n^R )  }$. 
This function represents the cascaded MA–TRIS–user channel obtained by summing the contributions of all $N$ TRIS elements. The phase term $\psi_n(\boldsymbol{t})$ denotes the effective phase of the $n$-th path, which consists of the applied TRIS phase shift $\tilde{\phi}_n$ and the propagation-induced phase shift determined by the total distance.

\begin{remark}
    To assess the impact of the MA, we evaluate (\ref{eq:SNRFirst}) for both discrete and continuous TRIS phase configurations. In the ideal case with continuous phase-shifters, each TRIS element can achieve exact phase alignment with the desired transmission direction, eliminating residual phase mismatch. In other words, for any value of $\frac{2\pi}{\lambda}(d_{n}^\text{T}(\boldsymbol{t})+d_{n}^\text{R})$, the phase $\phi_n$ can be chosen to perfectly compensate the propagation-induced phase. 
    In this scenario, coherent signal combining is fully achieved through phase control, and the received SNR reaches its maximum value. The resulting performance therefore serves as an upper bound for systems employing practical discrete phase-shifters.
    The MA position still affects the propagation distances, but perfect phase alignment ensures that the achievable SNR represents the theoretical upper bound.
    \begin{align} \label{eq:UpperBound}
        & \text{SNR}_\text{UB} = 
         \kappa \left | \sum_{n = 1}^{N}\frac{1}{d_{n}^\text{T}d_{n}^\text{R}}\right |^2
    \end{align}
\end{remark}

\subsection{Problem Formulation}
In this part, we focus on low-complexity TRIS implementations with discrete phase shifters.
The goal is now to maximize the SNR in (\ref{eq:SNRFirst}) by optimizing the MA location and the discrete phase-shifters of the TRIS. The optimization problem can be formulated as 
\begin{subequations}\label{eq:series}
    \begin{align}
        \mathcal{P}_1:\ \quad & \max_{\boldsymbol{t},\boldsymbol{\tilde{\phi}}} \quad \text{SNR}(\boldsymbol{t},\tilde{\boldsymbol{\phi}}) , \label{eq:SNR}\\     
        \text{s.t.}  \quad & \boldsymbol{t} \in \mathcal{R}_\text{MA},  \label{eq:MA.Reg}\\  
                     \quad & \lvert \tilde{\phi}_n \rvert = 1, && \forall n \in \mathcal{N} \label{eq:phiConstraint} \\
                     \quad & \arg ( \tilde{\phi}_n ) \in \boldsymbol{\Phi}_b, && \forall n \in \mathcal{N} \label{eq:MADiscrete}
    \end{align}
\end{subequations}
where \eqref{eq:MA.Reg} defines the two dimensional feasible region for antenna movement represented by $\mathcal{R}_\text{MA}$.
The \eqref{eq:phiConstraint} describes the constraint of the TRIS phase-shifters, and the \eqref{eq:MADiscrete} ensures that the discrete phase-shifters are inside the defined feasible set denoted by $\boldsymbol{\Phi}_b$. The optimization problem $\mathcal{P}_1$ is a non-convex problem due to the discrete phase constraints and the nonlinear coupling between $\boldsymbol{t}$ and $\tilde{\boldsymbol{\phi}}$.

\section{Proposed Algorithm} 
To solve the optimization problem $\mathcal{P}_1$, we employ an alternating optimization (AO) framework, which decomposes problem $\mathcal{P}_1$ into two subproblems: MA position optimization with respect to $\boldsymbol{t}$ and TRIS phase-shift optimization with respect to $\tilde{\boldsymbol{\phi}}$. Each variable is iteratively optimized while keeping the other fixed, and the two subproblems are solved alternately until convergence.

\begin{algorithm}[t]
\caption{AO Algorithm for Joint Optimization of $\boldsymbol{t}$ and $\tilde{\boldsymbol{\phi}}$}
\label{alg:AO}
\begin{algorithmic}[1]

\State \textbf{Initialize:} 
$\boldsymbol{t}^{(0)}$, 
$\tilde{\boldsymbol{\phi}}^{(0)}$, 
step size $\mu_0$, 
minimum step size $\mu_{\min}$, 
threshold $\epsilon$, 
outer index $i=0$, 
maximum outer iterations $i_{\max}$.

\State Compute $\mathrm{SNR}^{(0)} = \mathrm{SNR}(\boldsymbol{t}^{(0)},\tilde{\boldsymbol{\phi}}^{(0)})$.

\Repeat

    \State \textbf{MA position update (GA):}
    \State Set $q=0$ and $\boldsymbol{t}^{(i,q)}=\boldsymbol{t}^{(0)}$

    \Repeat
        \State Compute $\nabla_{\boldsymbol{t}}\mathrm{SNR}(\boldsymbol{t}^{(i,q)},\tilde{\boldsymbol{\phi}}^{(i)})$ using~(\ref{eq:grad})
        \State $\mu \leftarrow \mu_0$

        \Repeat
            \State $\hat{\boldsymbol{t}} = 
            \boldsymbol{t}^{(i,q)} 
            + \mu \nabla_{\boldsymbol{t}}\mathrm{SNR}
            (\boldsymbol{t}^{(i,q)},\tilde{\boldsymbol{\phi}}^{(i)})$
            \State $\mu \leftarrow \mu/2$
        \Until{\algcond{
        $(\ref{eq:MA.Reg})$ and 
        $\mathrm{SNR}(\hat{\boldsymbol{t}},\tilde{\boldsymbol{\phi}}^{(i)}) 
        > \mathrm{SNR}(\boldsymbol{t}^{(i,q)},\tilde{\boldsymbol{\phi}}^{(i)})$ \\
        or $\mu < \mu_{\min}$
        }}

        \State $\boldsymbol{t}^{(i,q+1)} \leftarrow \hat{\boldsymbol{t}}$
        \State $q \leftarrow q+1$

    \Until{$q=q_{\max}$} 

    \State $\boldsymbol{t}^{(i+1)} \leftarrow \boldsymbol{t}^{(i,q)}$

    \State \textbf{TRIS phase update (quantized alignment):}
    \For{$n=1,\ldots,N$}
        \State 
        $\tilde{\phi}_n^{(i+1)} =
        Q\!\left(
        \frac{2\pi}{\lambda}
        \big(d_n^{\mathrm{R}}(\boldsymbol{t}^{(i+1)}) + d_n^{\mathrm{T}}\big);
        \boldsymbol{\Phi}_b
        \right)$
    \EndFor

    \State 
    $\tilde{\boldsymbol{\phi}}^{(i+1)}
    \leftarrow
    [\tilde{\phi}_1^{(i+1)},\ldots,\tilde{\phi}_N^{(i+1)}]^T$

    \State Compute 
    $\mathrm{SNR}^{(i+1)} =
    \mathrm{SNR}(\boldsymbol{t}^{(i+1)},\tilde{\boldsymbol{\phi}}^{(i+1)})$

    \State $i \leftarrow i+1$

\Until{$|\mathrm{SNR}^{(i)}-\mathrm{SNR}^{(i-1)}|\le\epsilon$
or $i=i_{\max}$}

\State \textbf{Output:} 
$\boldsymbol{t}^\star=\boldsymbol{t}^{(i)}$, 
$\tilde{\boldsymbol{\phi}}^\star=\tilde{\boldsymbol{\phi}}^{(i)}$

\end{algorithmic}
\end{algorithm}

\subsection{Optimizing the MA position vector $\boldsymbol{t}$}
To determine the optimal location for the MA, we implement a gradient ascent (GA) approach. 
The calculated gradient value of $\text{SNR}(\boldsymbol{t},\tilde{\boldsymbol{\phi}})$ with respect to $\boldsymbol{t}$ is expressed as 
\begin{equation} \label{eq:grad}
\begin{split}
    & \nabla_{\boldsymbol{t}} \text{SNR}(\boldsymbol{t},\tilde{\boldsymbol{\phi}}) = \\ 
    & 2 \kappa \Re \Big\{ \Big( -\sum_{n=1}^{N} \frac{e^{j\psi(\boldsymbol{t})} (\boldsymbol{t}-\boldsymbol{e}_n) }{d^R_n d_n^T(\boldsymbol{t})^2} ( j\frac{2\pi}{\lambda} + \frac{1}{d_n^T(\boldsymbol{t})} )  \Big) c^*(\boldsymbol{t},\tilde{\boldsymbol{\phi}}) \Big\}.
\end{split}
\end{equation}

\subsection{Optimizing the TRIS phase-shift vector $\tilde{\boldsymbol{\phi}}$}
When employing a TRIS with discrete phase-shift adjustment, the limited resolution of the phase-shifters must be considered. 
We define $Q(\cdot)$ as the \emph{quantization function}, which projects the ideal phase-shift onto the nearest element of the feasible phase set $\boldsymbol{\Phi}_b$ associated with the TRIS phase-shifters
\begin{align}
     \tilde{{\phi}}_{n} = Q({\phi}_{n};\boldsymbol{\Phi}_b) = \arg\min_{{\phi}_b \in \boldsymbol{\Phi}_b} \, |{\phi}_{n} - {{\phi}_b}|.
\end{align} 
As a result, each element of the quantized phase-shifter vector $\tilde{\boldsymbol{\phi}}$ is calculated as 
\begin{align}\label{phi_Dis}
    \tilde{\phi}_{n} = Q \big( \frac{2\pi}{\lambda} (d_{n}^\text{R} + d_{n}^\text{T});\boldsymbol{\Phi}_b \big). 
\end{align}

\subsection{Convergence and Complexity Analysis}
The corresponding algorithm to optimize $\boldsymbol{\tilde{\phi}}$ and $\boldsymbol{t}$ is given in Algorithm \ref{alg:AO}. The algorithm follows an AO framework, where the MA position is updated via GA with backtracking and the TRIS phase vector is updated by element-wise quantized phase alignment. 
Due to the backtracking rule, each accepted MA update yields a non-decreasing SNR, while the TRIS phase update maximizes coherent combining under the discrete constraint. 
Since the SNR is upper bounded over the feasible MA region and the finite phase set, the sequence $\{\mathrm{SNR}^{(i)}\}$ is bounded and convergent. 
However, due to the non-convex nature of the problem, the algorithm converges in general to a local stationary solution. 
Regarding complexity, each gradient computation and SNR evaluation requires $\mathcal{O}(N)$ operations, where $N$ is the number of TRIS elements. 
Hence, the overall complexity scales linearly with $N$, i.e., $\mathcal{O}(i_{\max} q_{\max} N)$.

\section{Simulation Results}
In this section, simulations are presented to illustrate the effects of employing a MA under a discrete phase-shifter TRIS condition.
We assume the carrier frequency $f = 20\,\mathrm{GHz}$ and TRIS element spacing of $\lambda/2$ and the noise variance $\sigma^2 =~-70\,\mathrm{dBm}$. The user is located at a distance of $50\, \mathrm{m}$ of the TRIS. 

\begin{figure}[ht]
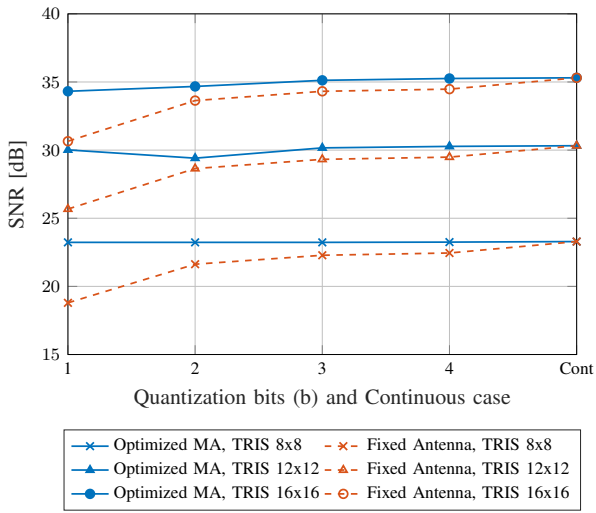

  \centering
  \includestandalone[width=0.9\linewidth]{Figures/SNR_vs_Bits} 
  \caption{SNR vs. quantization levels for optimized MA and fixed position antenna.}
  \label{fig:snr_vs_bits}
\end{figure}
Fig.~\ref{fig:snr_vs_bits} illustrates the impact of phase quantization resolution on the achievable SNR for different TRIS sizes. We consider a transmit power of $P = 13.6\,\mathrm{dBm}$ and a distance of $0.5\,\mathrm{m}$ between the MA and the TRIS. As the number of quantization bits $b$ increases, the SNR improves and gradually approaches the continuous-phase upper bound (cont. case at right edge of the plot), since a higher phase resolution enables more accurate phase alignment across the TRIS elements. 
Furthermore, optimizing the MA position consistently provides an SNR gain over the fixed-antenna case for all quantization levels and TRIS sizes. This shows that the additional spatial degree of freedom introduced by the MA can partially mitigate the performance degradation caused by phase quantization. However, the remaining gap to the continuous-phase benchmark indicates that MA repositioning cannot completely remove the quantization loss. It can also be observed that this mitigation effect is relatively more pronounced for smaller TRIS sizes. In this case, the quantization-induced mismatch is less severe and can be more effectively compensated by adjusting the MA location. For larger TRIS apertures, the discrete phase errors accumulate over more transmissive elements, making the performance increasingly sensitive to phase resolution and reducing the relative compensation achievable by a single MA.

\begin{figure}[ht]
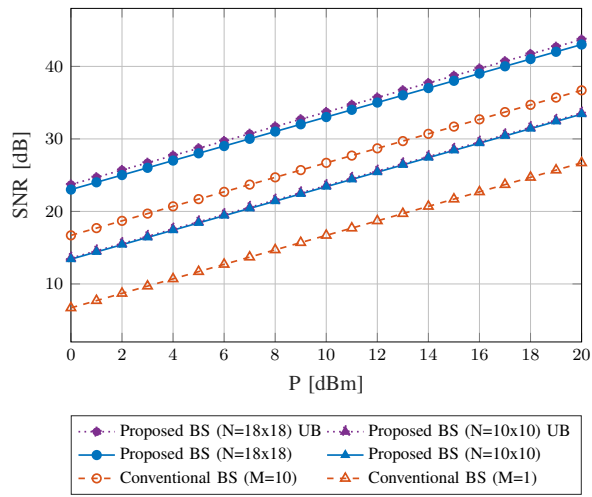

  \centering
  \includestandalone[width=0.9\linewidth]{Figures/SNR_vs_Pmax} 
  \caption{Comparison of the SNR performance for the proposed BS configuration with a conventional BS.}
  \label{fig:compact_vs_conventional}
\end{figure}
Moreover, Fig.~\ref{fig:compact_vs_conventional} compares the SNR performance of the proposed BS architecture with a conventional multiple antenna BS as a function of the transmit power constraint $P$. 
As a benchmark, we consider a conventional fully-active BS equipped with a fixed antenna array with $M \in \{1, 10\}$ number of antennas. All antenna elements are connected to dedicated RF chains and perform beamforming using maximum-ratio transmission (MRT) under the same transmit power constraint and channel model as the proposed architecture. This baseline represents the typical active antenna array employed in current wireless systems.
The proposed BS consists of one MA combined with a TRIS with 2 bits quantization level and different sizes of $N \in \{10\times10, 18\times18\}$. The distance between the MA and the TRIS is $0.5\,\mathrm{m}$. The case with continuous-phase TRIS elements is also shown as the upper bound. As expected, the SNR increases approximately linearly (in dB scale) with $P$ for all configurations. The proposed BS consistently outperforms the conventional counterpart across the entire power range.
The performance gain stems from the large effective aperture provided by the TRIS and the additional spatial degree of freedom introduced by MA repositioning. Even though the proposed BS employs only a single active antenna, the passive TRIS enables coherent signal combining across a large number of elements, which can provide a higher array gain than conventional active antenna arrays. 

\begin{figure}[ht]
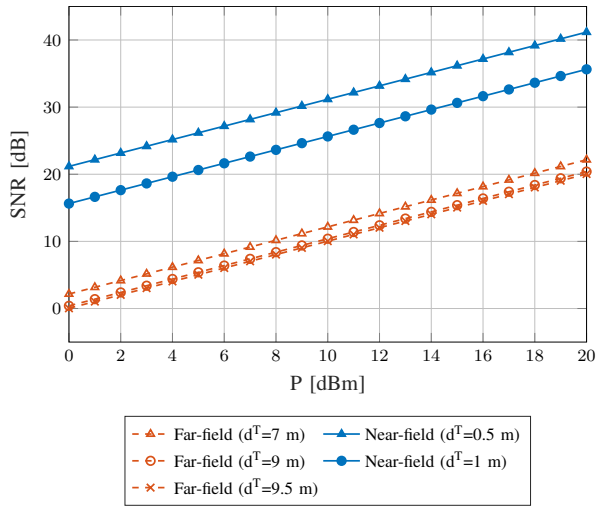

  \centering
  \includestandalone[width=0.9\linewidth]{Figures/NF_FF_TRIS_vs_Pmax} 
  \caption{SNR versus $P$ for near-field and far-field TRIS placements.}
  \label{fig:FF_NF_TRIS}
\end{figure}
Additionally, Fig.~\ref{fig:FF_NF_TRIS} illustrates the SNR versus $P$ for the proposed BS architecture under near-field and far-field configurations while enforcing a constant total propagation distance $d_{\text{Tot}}=d^T+d^R=50\,\mathrm{m}$ between the MA and the user, while varying the distance between the MA and the TRIS. Despite the identical total distance, a clear performance gap between the near-field and far-field regimes is observed due to the cascaded path-loss structure of the MA–TRIS–user link. Unlike direct transmission, where the received power scales with $1/d_{\text{Tot}}^2$, the TRIS-assisted link approximately follows $\left(\sum_{n} 1/(d_n^T d_n^R)\right)^2$, resulting in product-type attenuation. In the near-field regime, the distances from the MA to each $n$-th TRIS element $(d_n^T)$ vary significantly across the TRIS aperture, and some elements experience shorter propagation distances. This distance diversity improves coherent signal combining and partially mitigates the cascaded path loss. Moreover, near-field propagation enables highly localized spatial focusing, making the effective channel gain sensitive to small geometric variations between the MA and TRIS elements. In contrast, in the far-field regime the element-wise distances become nearly uniform, which reduces this effect and limits the achievable array gain. As shown in Fig.~\ref{fig:FF_NF_TRIS}, increasing the MA–TRIS distance therefore leads to a noticeable SNR degradation due to the cascaded path-loss structure of the TRIS-assisted link. Consequently, the performance advantage of the proposed architecture over a conventional fully-active BS is expected to decrease when the MA–TRIS separation becomes large, since the cascaded attenuation becomes more dominant. However, in the considered regime where the TRIS is deployed close to the transmitter, the proposed architecture consistently maintains a clear SNR advantage.

\section{Conclusion}
In this paper, we proposed a BS architecture that integrates a MA with a TRIS operating in near-field distance. By jointly optimizing the MA position and the discrete TRIS phase-shifters through an AO framework, significant SNR gains are achieved compared to conventional BS designs. Simulation results show that the MA--TRIS architecture consistently outperforms fully-active antenna arrays under the same transmit power constraint. Furthermore, increasing the TRIS aperture size and phase resolution improves coherent combining gain and approaches the continuous-phase upper bound. 
These results highlight the potential of exploiting near-field propagation dynamics and antenna mobility to enhance discrete-phase TRIS-assisted wireless communication systems.

\FloatBarrier
\bibliographystyle{IEEEtran}
\bibliography{Bib.bib}

\end{document}

%% file: Figures/SNR_vs_Bits.tex
\definecolor{mycolor1}{rgb}{0.00000,0.44700,0.74100}%
\definecolor{mycolor2}{rgb}{0.85000,0.32500,0.09800}%
\definecolor{mycolor3}{rgb}{0.92900,0.69400,0.12500}%
\begin{tikzpicture}

\begin{axis}[%
width=8.207cm,
height=5.5cm,
at={(0cm,0cm)},
scale only axis,
xmin=1,
xmax=5,
xtick={1,2,3,4,5},
xticklabels={{1},{2},{3},{4},{Cont}},
xlabel style={font=\color{white!15!black}},
xlabel={Quantization bits (b) and Continuous case},
ymin=15,
ymax=40,
ylabel style={font=\color{white!15!black}},
ylabel={$\text{SNR [dB]}$}, 
axis background/.style={fill=white},
title style={font=\bfseries},
xmajorgrids,
ymajorgrids,
legend style={
	at={(0.5,-0.22)},        
	anchor=north,
	legend columns=2,        
	legend cell align=left,
	draw=white!15!black
}
]
\addplot [color=mycolor1, line width=0.8pt, mark=x, mark options={solid, mycolor1}, mark size = 2.5pt]
  table[row sep=crcr]{%
1	23.2306180993174\\
2	23.2306181025882\\
3	23.2306181006065\\
4	23.2490515860413\\
5	23.288719150632\\
};
\addlegendentry{Optimized MA, TRIS 8x8} 

\addplot [color=mycolor2, dashed, line width=0.8pt, mark=x, mark options={solid, mycolor2}, mark size = 2.5pt]
  table[row sep=crcr]{%
1	18.7929543730021\\
2	21.6273003292686\\
3	22.2842319261619\\
4	22.4506763706323\\
5	23.288719150632\\
};
\addlegendentry{Fixed Antenna, TRIS 8x8} 

\addplot [color=mycolor1, line width=0.8pt, mark=triangle*, mark options={solid, mycolor1}]
  table[row sep=crcr]{%
1	30.0151256683885\\
2	29.4091155738557\\
3	30.1640297362134\\
4	30.2730165064876\\
5	30.3226428291504\\
};
\addlegendentry{Optimized MA, TRIS 12x12} 

\addplot [color=mycolor2, dashed, line width=0.8pt, mark=triangle, mark options={solid, mycolor2}]
  table[row sep=crcr]{%
1	25.6751428036041\\
2	28.6460594880123\\
3	29.3226079090878\\
4	29.4858108826987\\
5	30.3226428291504\\
};
\addlegendentry{Fixed Antenna, TRIS 12x12} 

\addplot [color=mycolor1, line width=0.8pt, mark=*, mark options={solid, mycolor1}]
  table[row sep=crcr]{%
1	34.3175456456607\\
2	34.6725184202962\\
3	35.1217355365167\\
4	35.2556750450738\\
5	35.3066330883348\\
};
\addlegendentry{Optimized MA, TRIS 16x16} 

\addplot [color=mycolor2, dashed, line width=0.8pt, mark=o, mark options={solid, mycolor2}]
  table[row sep=crcr]{%
1	30.6510704417581\\
2	33.6356803206689\\
3	34.3159578268037\\
4	34.4756584321076\\
5	35.3066330883348\\
};
\addlegendentry{Fixed Antenna, TRIS 16x16} 

\end{axis}
\end{tikzpicture}%

%% file: Figures/SNR_vs_Pmax.tex
	
\definecolor{mycolor1}{rgb}{0.00000,0.44700,0.74100}%
\definecolor{mycolor2}{rgb}{0.85000,0.32500,0.09800}%
\definecolor{mycolor3}{rgb}{0.92900,0.69400,0.12500}%
\definecolor{mycolor4}{rgb}{0.49400,0.18400,0.55600}%
\definecolor{mycolor5}{rgb}{0.46600,0.67400,0.18800}%
\definecolor{mycolor6}{rgb}{0.30100,0.74500,0.93300}%
\definecolor{mycolor7}{rgb}{0.63500,0.07800,0.18400}%

\definecolor{deepPurple}{RGB}{102,0,153}

\begin{tikzpicture}

\begin{axis}[%
width=8.414cm,
height=5.5cm,
at={(0cm,0cm)},
scale only axis,
xmin=0,
xmax=20,
xlabel style={font=\color{white!15!black}},
xlabel={$\text{P}\text{ [dBm]}$},
ymin=2,
ymax=48,
ylabel style={font=\color{white!15!black}},
ylabel={SNR [dB]},
axis background/.style={fill=white},
title style={font=\bfseries},
xmajorgrids,
ymajorgrids,
legend style={
    at={(0.5,-0.22)},        
    anchor=north,
    legend columns=2,        
    legend cell align=left,
    draw=white!15!black
}
]

\addplot [dotted, color=mycolor4, mark = *, line width=0.8pt]
  table[row sep=crcr]{%
0	23.706974576834\\
1	24.706974576834\\
2	25.706974576834\\
3	26.706974576834\\
4	27.706974576834\\
5	28.706974576834\\
6	29.706974576834\\
7	30.706974576834\\
8	31.706974576834\\
9	32.706974576834\\
10	33.706974576834\\
11	34.706974576834\\
12	35.706974576834\\
13	36.706974576834\\
14	37.706974576834\\
15	38.706974576834\\
16	39.706974576834\\
17	40.706974576834\\
18	41.706974576834\\
19	42.706974576834\\
20	43.706974576834\\
};
\addlegendentry{$\text{Proposed BS (N=18x18) UB}$}

\addplot [dotted, color=mycolor4, line width=0.8pt, mark=triangle*, mark size = 3 pt]
  table[row sep=crcr]{%
0	13.5607509864297\\
1	14.5607509864297\\
2	15.5607509864297\\
3	16.5607509864297\\
4	17.5607509864297\\
5	18.5607509864297\\
6	19.5607509864297\\
7	20.5607509864297\\
8	21.5607509864297\\
9	22.5607509864297\\
10	23.5607509864297\\
11	24.5607509864297\\
12	25.5607509864297\\
13	26.5607509864297\\
14	27.5607509864297\\
15	28.5607509864297\\
16	29.5607509864297\\
17	30.5607509864297\\
18	31.5607509864297\\
19	32.5607509864297\\
20	33.5607509864297\\
};
\addlegendentry{$\text{Proposed BS (N=10x10) UB}$}

\addplot [solid, mark=*, color=mycolor1, line width=0.8pt]
table[row sep=crcr]{%
0	23.0153353968054\\
1	24.0153353968054\\
2	25.0153353968054\\
3	26.0153353968054\\
4	27.0153353968054\\
5	28.0153353968054\\
6	29.0153353968054\\
7	30.0153353968054\\
8	31.0153353968054\\
9	32.0153353968054\\
10	33.0153353968054\\
11	34.0153353968054\\
12	35.0153353968054\\
13	36.0153353968054\\
14	37.0153353968054\\
15	38.0153353968054\\
16	39.0153353968054\\
17	40.0153353968054\\
18	41.0153353968054\\
19	42.0153353968054\\
20	43.0153353968054\\
};
\addlegendentry{Proposed BS (N=18x18)}

\addplot [solid, mark=triangle*, color=mycolor1, line width=0.8pt ]
table[row sep=crcr]{%
0	13.4147602858211\\
1	14.4147602858211\\
2	15.4147602858211\\
3	16.4147602858211\\
4	17.4147602858211\\
5	18.4147602858211\\
6	19.4147602858211\\
7	20.4147602858211\\
8	21.4147602858211\\
9	22.4147602858211\\
10	23.4147602858211\\
11	24.4147602858211\\
12	25.4147602858211\\
13	26.4147602858211\\
14	27.4147602858211\\
15	28.4147602858211\\
16	29.4147602858211\\
17	30.4147602858211\\
18	31.4147602858211\\
19	32.4147602858211\\
20	33.4147602858211\\
};
\addlegendentry{Proposed BS (N=10x10)}


\addplot [dashed, mark=o, mark options={solid}, color=mycolor2, line width=0.8pt]
table[row sep=crcr]{%
0	16.6999800314784\\
1	17.6999800314784\\
2	18.6999800314784\\
3	19.6999800314784\\
4	20.6999800314784\\
5	21.6999800314784\\
6	22.6999800314784\\
7	23.6999800314784\\
8	24.6999800314784\\
9	25.6999800314784\\
10	26.6999800314784\\
11	27.6999800314784\\
12	28.6999800314784\\
13	29.6999800314784\\
14	30.6999800314784\\
15	31.6999800314784\\
16	32.6999800314784\\
17	33.6999800314784\\
18	34.6999800314784\\
19	35.6999800314784\\
20	36.6999800314784\\
};
\addlegendentry{Conventional BS (M=10)}

\addplot [dashed, mark=triangle, mark options={solid}, color=mycolor2, line width=0.8pt, mark size = 2.5pt]
  table[row sep=crcr]{%
0	6.69998082065976\\
1	7.69998082065976\\
2	8.69998082065976\\
3	9.69998082065976\\
4	10.6999808206598\\
5	11.6999808206598\\
6	12.6999808206598\\
7	13.6999808206598\\
8	14.6999808206598\\
9	15.6999808206598\\
10	16.6999808206598\\
11	17.6999808206598\\
12	18.6999808206598\\
13	19.6999808206598\\
14	20.6999808206598\\
15	21.6999808206598\\
16	22.6999808206598\\
17	23.6999808206598\\
18	24.6999808206598\\
19	25.6999808206598\\
20	26.6999808206598\\
};
\addlegendentry{Conventional BS (M=1)}

\end{axis}
\end{tikzpicture}%

%% file: Figures/NF_FF_TRIS_vs_Pmax.tex
\definecolor{mycolor1}{rgb}{0.00000,0.44700,0.74100}%
\definecolor{mycolor2}{rgb}{0.85000,0.32500,0.09800}%
\begin{tikzpicture}

\begin{axis}[%
width=8.475cm,
height=5.5cm,
at={(0cm,0cm)},
scale only axis,
xmin=0,
xmax=20,
xlabel style={font=\color{white!15!black}},
xlabel={$\text{P}\text{ [dBm]}$},
ymin=-5,
ymax=45,
ylabel style={font=\color{white!15!black}},
ylabel={SNR [dB]},
axis background/.style={fill=white},
xmajorgrids,
ymajorgrids,
legend style={
	at={(0.5,-0.22)},        
	anchor=north,
	legend columns=2,        
	legend cell align=left,
	draw=white!15!black
}
]

\addplot [color=mycolor2, dashed, line width=0.8pt, mark=triangle, mark options={solid, mycolor2}]
  table[row sep=crcr]{%
0	2.16022315935437\\
1	3.16022315935437\\
2	4.16022315935437\\
3	5.16022315935437\\
4	6.16022315935437\\
5	7.16022315935437\\
6	8.16022315935437\\
7	9.16022315935437\\
8	10.1602231593544\\
9	11.1602231593544\\
10	12.1602231593544\\
11	13.1602231593544\\
12	14.1602231593544\\
13	15.1602231593544\\
14	16.1602231593544\\
15	17.1602231593544\\
16	18.1602231593544\\
17	19.1602231593544\\
18	20.1602231593544\\
19	21.1602231593544\\
20	22.1602231593544\\
};
\addlegendentry{$\text{Far-field  (d}^\text{T}\text{=7 m)}$}

\addplot [color=mycolor1, line width=0.8pt, mark=triangle*, mark options={solid, mycolor1}]
  table[row sep=crcr]{%
0	21.1683648701483\\
1	22.1683648701483\\
2	23.1683648701483\\
3	24.1683648701483\\
4	25.1683648701483\\
5	26.1683648701483\\
6	27.1683648701483\\
7	28.1683648701483\\
8	29.1683648701483\\
9	30.1683648701483\\
10	31.1683648701483\\
11	32.1683648701483\\
12	33.1683648701483\\
13	34.1683648701483\\
14	35.1683648701483\\
15	36.1683648701483\\
16	37.1683648701483\\
17	38.1683648701483\\
18	39.1683648701483\\
19	40.1683648701483\\
20	41.1683648701483\\
};
\addlegendentry{$\text{Near-field (d}^\text{T}\text{=0.5 m)}$}

\addplot [color=mycolor2, dashed, line width=0.8pt, mark=o, mark options={solid, mycolor2}]
  table[row sep=crcr]{%
0	0.394704248979947\\
1	1.39470424897995\\
2	2.39470424897995\\
3	3.39470424897995\\
4	4.39470424897995\\
5	5.39470424897995\\
6	6.39470424897995\\
7	7.39470424897995\\
8	8.39470424897995\\
9	9.39470424897995\\
10	10.3947042489799\\
11	11.3947042489799\\
12	12.3947042489799\\
13	13.3947042489799\\
14	14.3947042489799\\
15	15.3947042489799\\
16	16.3947042489799\\
17	17.3947042489799\\
18	18.3947042489799\\
19	19.3947042489799\\
20	20.3947042489799\\
};
\addlegendentry{$\text{Far-field  (d}^\text{T}\text{=9 m)}$} 

\addplot [color=mycolor1, line width=0.8pt, mark=*, mark options={solid, mycolor1}]
  table[row sep=crcr]{%
0	15.6323379119983\\
1	16.6323379119983\\
2	17.6323379119983\\
3	18.6323379119983\\
4	19.6323379119983\\
5	20.6323379119983\\
6	21.6323379119983\\
7	22.6323379119983\\
8	23.6323379119983\\
9	24.6323379119983\\
10	25.6323379119983\\
11	26.6323379119983\\
12	27.6323379119983\\
13	28.6323379119983\\
14	29.6323379119983\\
15	30.6323379119983\\
16	31.6323379119983\\
17	32.6323379119983\\
18	33.6323379119983\\
19	34.6323379119983\\
20	35.6323379119983\\
};
\addlegendentry{$\text{Near-field (d}^\text{T}\text{=1 m)}$}

\addplot [color=mycolor2, dashed, line width=0.8pt, mark=x, mark options={solid, mycolor2}, mark size = 2.5pt]
  table[row sep=crcr]{%
0	0.0322453720843153\\
1	1.03224537208432\\
2	2.03224537208432\\
3	3.03224537208431\\
4	4.03224537208431\\
5	5.03224537208432\\
6	6.03224537208432\\
7	7.03224537208432\\
8	8.03224537208431\\
9	9.03224537208431\\
10	10.0322453720843\\
11	11.0322453720843\\
12	12.0322453720843\\
13	13.0322453720843\\
14	14.0322453720843\\
15	15.0322453720843\\
16	16.0322453720843\\
17	17.0322453720843\\
18	18.0322453720843\\
19	19.0322453720843\\
20	20.0322453720843\\
};
\addlegendentry{$\text{Far-field  (d}^\text{T}\text{=9.5 m)}$}

\end{axis}
\end{tikzpicture}